\documentstyle[preprint,aps,epsfig]{revtex}

\begin{document}

\preprint{ECT$^\ast$-01-041; UTF/446}             

\draft

\title{Spin force dependence of the parton distributions: \\
the ratio $F_2^n(x,Q^2)/F_2^p(x,Q^2)$}
\author{Barbara Pasquini$^{1,3}$, Marco Traini$^{2,3}$, and Sigfrido Boffi$^{4,5}$}
\address{
$^1$ ECT$^\ast$, Strada delle Tabarelle 286,
I-38050 Villazzano (Trento), Italy \\
$^2$ Dipartimento di Fisica, Universit\`a degli Studi di Trento,
I-38050 Povo (Trento), Italy\\
$^3$ INFN, Trento, Italy \\
$^4$Dipartimento di Fisica Nucleare e Teorica,
Universit\`a degli Studi di Pavia, I-27100 Pavia, Italy \\
$^5$ INFN, Pavia, Italy}

\maketitle


\begin{abstract}
Light-front Hamiltonian dynamics is used to relate low-energy constituent 
quark models to deep inelastic unpolarized structure functions of the nucleon.
 The approach incorporates the correct Pauli principle prescription  
consistently and it allows a transparent investigation of the effects due to 
the spin-dependent $SU(6)$-breaking terms in the quark model Hamiltonian. Both
 Goldstone-boson-exchange interaction and hyperfine-potential models are 
discussed in a unified scheme and a detailed comparison, between the two 
(apparently) different potential prescriptions, is presented.

\end{abstract}


\newpage


\section{Introduction}

Constituent quark models (CQM) provide a basic tool for the 
description of low-energy hadron phenomena. Important features
of non-perturbative quantum chromodynamics (QCD) can be incorporated
in the CQM providing a framework for quantitative calculations of
hadron properties and reaction observables. In particular
CQM can be made to incorporate some of the basic symmetries 
\cite{Morp}
of QCD and, at the same time, can be defined within relativistic 
frameworks where covariance is preserved \cite{HamDy}.

Gluons and quark-antiquark pairs surrounding a current quark are 
considered an integral part of it and together they make the 
constituent quark: a picture which seems to be 
substantiated by recent lattice QCD results \cite{lat-CQM}.

Once explicit gluon degrees of freedom are integrated out, spin-dependent 
forces emerge within a potential description. Such forces 
between quarks have been associated to the 
chromomagnetic interaction of one-gluon exchange (OGE) \cite{oge}, 
to Goldstone-boson exchange (GBE) in connection with the 
breaking of chiral symmetry \cite{gbe}, and to the effects of 
instantons \cite{instan}. At present all three models 
seem to have enough flexibility to approximate, to a large extent, existing 
data and the question whether they are just different ways of describing the 
same thing, remains unanswered. Actually the debate on the ``defects'' of the 
OGE as well as of the GBE is particularly active \cite{isgur00nt,glo00nt}. 
The investigation has been mainly 
confined to low-energy hadron excitation spectrum \cite{spectrum} and to the 
electroweak form factors of the nucleon \cite{static}. 
In the present work we want to enlarge considerably the front of 
a possible comparison between the OGE and the
GBE dynamical mechanisms considering the effects of the spin-dependent
forces on the nucleon deep inelastic responses (structure functions).

Despite the fact that the subject has a rather long story,
we can show that only at present we have the necessary tools
to address the problem in a simple and transparent way and to elucidate the 
important role of spin-dependent forces on 
parton distributions. Let us begin summarizing existing approaches.

F.E. Close in 1973 \cite{close73} discussed the sensitivity of the ratio  
$F_2^n/F_2^p$ in relation with spin-dependent forces. He used a model of the 
nucleon where the nucleon first breaks up into a quark (which  then interacts 
with the electromagnetic field) and a ``quasi'' particle called core. The core 
can have spin 0 or 1 and these two components have equal probability in a 
$SU(6)$ symmetric model only (where spin-dependent forces are neglected). He 
concluded that the predictions for the ratio $F_2^n/F_2^p$ are related to the 
high-momentum behaviour of the proton-$\Delta$ transition form factors.

Carlitz \cite{carlitz75} investigated the $SU(6)$ symmetry breaking effects 
in deep inelastic scattering concluding that the mass difference between
$N$ and $\Delta$ states of the {\underline {56}} baryons 
implies that the ratio $F_2^n/F_2^p$  should approach 1/4 as $x \to 1$.
In a more recent work Close and Thomas \cite{closethomas88} related 
the $\Delta$-N mass difference, to the 
ratio of neutron and proton inelastic structure functions and to deep 
inelastic polarization asymmetries.

Nathan Isgur \cite{isgur99} in a quite recent paper investigates the
hyperfine-perturbed quark model to make predictions for the nucleon
spin-dependent distribution functions and shows that precise
measurements of the asymmetries $A_1^p(x)$ and $A_1^n(x)$ (in the 
valence region) can test the model and verify the ``normal'' behaviour
of the valence-quark spin distribution at variance with the (sometimes) invoked
 ``spin crisis''. In the same paper Isgur critically
discusses the history of the $SU(6)$-breaking effects with
particular emphasis on the ratio $F_2^n/F_2^p$. In fact both papers of 
Close and Carlitz and Thomas studied the behaviour of the 
spin 1 ($\psi^1$) and 
zero ($\psi^0$) pair wave-functions separately; an approximated scheme which is
 not consistent with the Pauli principle unless $\psi^1$ and $\psi^0$ have 
very special properties under the permutation group in three dimensions. Isgur 
concludes that the situation remains rather unclear also because most authors 
have attempted ``absolute'' calculations of the structure functions with the 
unavoidable need of assumptions, approximations and ``procedures''. 
A criticism which involves also other calculations \cite{rec1,rec2,rec3,weber94}.

Aim of the present paper is a close investigation of the relation between the 
$SU(6)$-breaking effects, responsible for the $\Delta$-N mass splitting and 
the charge radius of the neutron, and the ratio $F_2^n/F_2^p$.  In our 
approach the Pauli principle is fully satisfied and we can compare effects of 
spin-dependent forces originating from both: chromomagnetic (hyperfine) OGE 
interaction and chiral symmetry motivated GBE mechanism.
Because of the large debate on the subject, this last result is probably the 
most motivating one, and we show that the Pauli principle effect reveals to be
 of crucial importance in the discussion.

\section{Partons and quarks}

Let us introduce the method we use to calculate the 
proton and neutron structure functions and its quite natural connection
with the nucleon wave-function built in a given quark model.

The work by Gl\"uck, Reya and coworkers \cite{GRetal}
has shown that, starting from a parton parametrization at a low
resolution scale $\mu_0^2$, the experimental deep inelastic structure
functions at high-momentum transfer can be reproduced and even predicted
\cite{GRprediction}. $\mu_0^2$ is evaluated by evolving back the second 
moment of the valence distribution to the point where it becomes dominant.
The procedure, closely related to a suggestion due to Parisi and 
Petronzio \cite{PaPetr}, assumes
that there exists a scale, $\mu_0^2$, where the short range (perturbative)
part of the interaction is negligible, therefore the glue and sea are
suppressed, and the long range (confining) part of the interaction
produces a proton composed of three (valence)
quarks only. Jaffe and Ross \cite{jaffeross80} proposed thereafter to
ascribe the quark model calculations of matrix elements to the hadronic scale 
$\mu_0^2$. For larger $Q^2$ their Wilson coefficients will give the evolution 
as dictated by perturbative QCD. In this way quark models, summarizing a great 
deal of hadronic properties,
may substitute low-energy parametrizations\footnote{
Let us stress that for the QCD evolution one needs matrix elements
of the twist-two part of the current and not the calculation of the full 
response at low scale.}.

Following such a path, a partonic description can be generated from gluon
radiation even off a pure valence quark system, which can be used
to evaluate the non-perturbative input 
occurring in the Operator Product Expansion (OPE) analysis of 
lepton-hadron scattering in QCD \cite{evolving}.  A systematic analysis
shows that the approach can consistently be developed at 
Next-to-Leading Order both for polarized and unpolarized structure 
functions, including non-perturbative contributions from the nucleon 
cloud \cite{noi1} or from the partonic substructure of the constituent 
quarks\cite{SVT97ss}.  In addition it can consistently be improved
including relativistic covariance effects within the light-front Hamiltonian 
dynamics \cite{FTVlf99}.

In the light-front description of deep inelastic scattering,  the parton model 
is recovered, in the Bjorken limit, due to the dominance of short light-cone 
distances 
in the relevant Feynmann diagrams. As a consequence 
the partonic description can be developed in the rest frame of the hadron by 
using light-cone formalism.  In particular the $i$-th parton distribution can 
be related to the light-cone momentum density\footnote{A
formal derivation of Eq.~(\ref{DVal})  can be found in the paper by
Mair and Traini in ref.\cite{noi1}.}
\begin{equation} 
q_{i}^{\uparrow\,(\downarrow)}(x,\mu_0^2) = {1 \over (1-x)^2}\, \int
d^3k\,\,n_{i}^{\uparrow\,(\downarrow)}({\bf k}^2)\,\delta \left({x
\over 1-x} - {k^+\over M_N}\right)\,\,, \label{DVal} 
\end{equation} 
where $k^+/P^+=k^+/M_N=(\sqrt{{\bf k}^2+m_i^2}+k_z)/M_N$ is the light-cone
momentum fraction of the struck parton in the rest frame, $M_N$ and $m_i$
are the nucleon and parton mass respectively and $n_{i}^\uparrow({\bf
k}^2)$, $n_{i}^\downarrow({\bf k}^2)$ represent the light-cone momentum density
of the $i$-th parton whose spin is {\it aligned}
($\uparrow$) or {\it anti-aligned} ($\downarrow$) to the total spin of the
parent nucleon. If one assumes that at the scale $\mu_0^2$ only the $u$
and $d$ constituent quarks are resolved, the momentum densities can be
written 
\begin{equation} 
n^{\uparrow\,(\downarrow)}_{u(d)}({\bf k}^2)=\langle
N,J_z=+{1\over 2}|\sum_{i=1}^3\, {1+(-)\tau_i^z \over 2}\,{1+(-)\sigma_i^z 
\over
2}\, \delta({\bf k}-{\bf k}_i)|N,J_z=+{1\over 2}\rangle\,\, .  
\label{n_up/down}
\end{equation} 
The light-cone distributions (\ref{n_up/down}) can be
evaluated including relativistic effects as introduced by a light-front 
formulation of a three-body interacting system \cite{FTVlf99}. 
As a consequence we remain within a constituent picture where the partons in 
the rest frame are identified with three (constituent) quarks at the hadronic 
scale, and covariance requirement as well as Pauli principle are fulfilled.

Since the hadronic scale $\mu_0^2$ turns out to be very low  ($\mu_0^2 \sim 
(0.1 - 0.2)$ GeV$^2$), close to the constituent quark
mass\footnote{The 
actual value of the scale $\mu_0^2$ ranges from 0.1 GeV$^2$, if only valence 
quarks are considered, to 0.37 GeV$^2 \cite{noi1}$,
when non-perturbative $q-\bar q$ pairs and gluons are included.}, we
assume that the constituent picture at this scale 
is represented by a constituent quark model, 
with parameters fixed to reproduce the basic features of the nucleon spectrum 
in the
energy region $1 - 2$ GeV. The constituent quark models we
will make use of, in the present paper, are the Isgur-Karl model (IK)
proposed long ago \cite{IKmodel} and the GBE 
recently developed \cite{spectrum,static,grazprivate}.

In particular the unpolarized valence parton distributions
$
q^\uparrow(x,\mu_0^2)+q^\downarrow(x,\mu_0^2) = q_V(x,\mu_0^2),
$
(with $q\equiv u\,,d$), at the hadronic scale are related to the
scalar momentum densities $n_{u(d)}({\bf k})^2$ (with
$\int d^3{\bf k}\, n_{u(d)}({\bf k}^2) = 2(1)$) calculated making
use of the wave-functions of the IK or GBE models with no free
parameters. The dynamical effects due to the $SU(6)$-breaking terms
in the quark Hamiltonian are entirely embedded in the momentum 
densities and no approximation is required. Details of the approach 
can be found in ref.\cite{FTVlf99}.

\section{Results and discussion}

The results for the (twist-two part of) proton and neutron structure 
functions predicted by the two models, are shown in Figs.~1 both at 
the hadronic scale and experimental scale $Q^2 = 10$ GeV$^2$.
The $x$-dependence of the IK and GBE models differs quite substantially
in the large $x$-region both at the hadronic and at the experimental scale.
In particular the IK model lacks high-$x$ components: a result related
to the lack of high-momentum components in the nucleon wave-function. 
The relativized nature of the GBE Hamiltonian reflects into a larger
component of high momenta in the corresponding densities, an important 
ingredient to reproduce the behaviour of the structure functions at high-$x$
related to the presence of the $\sum_{i=1}^3 \sqrt{{\bf k}_i^2 + m_i^2}$
term in the Hamiltonian and therefore manifested in all the relativized models.

Before entering the discussion let us notice that the expected results for
$F_2^n$ and $F_2^p$ can be separated in two main regions. The low-$x$ 
($x\lesssim 0.3$) region will be dominated by gluonic and sea partons.
In the present calculation they are generated via bremsstrahlung radiation 
through renormalization group evolution. It is well known\cite{noi1} 
that the inclusion of these hard partons is not sufficient to explain the 
absolute values of the structure functions at low-$x$ and soft components 
have to be added at the hadronic scale to approach the data. 
Therefore we do not consider our calculation to be quantitative for that
region. At most we will obtain a qualitative description of the two structure
functions.

In the region of large-$x$ ($0.3 \lesssim x$) the structure functions  
$F_2^n$ and $F_2^p$ are dominated by valence parton effects which are 
sensitive to the quark model wave-functions. 
It is just that region where the results for the ratio $F_2^n/F_2^p$ 
can become a specific test for the model wave-functions.

\noindent Our main results are presented in Figs.~2, and a few comments are in 
order:

\noindent i) 
Despite of the large sea and gluon contributions produced by QCD evolution 
(as illustrated in the Figs.~1~a and 1~b) the ratio $F_2^n/F_2^p$ 
is scarcely influenced by the perturbative QCD radiative effects,
as expected.
$F_2^n(x,\mu_0^2)/F_2^p(x,\mu_0^2) \approx F_2^n(x,Q^2=10\,{\rm GeV}^2)/
F_2^p(x,Q^2=10\,{\rm GeV}^2)$ for the whole range 
$0.3 \lesssim x \lesssim 0.7$. Such insensitivity is, in our case, an
advantage. In fact the effects of QCD evolution largely cancel in the
ratio $F_2^n/F_2^p$ compensating also the uncertainties coming from
an evolution which starts from a quite low hadronic scale\footnote{
We do not need to enter a large discussion to justify our evolution 
approach because of such large cancellation. Details and discussions 
can be found in ref.\cite{noi1}.}. 
As a result, the details connected with the model wave-functions are 
emphasized (see Figs.~2~a,2~b).

\noindent ii) In particular it is evident that the Leading Order and 
the Next-to-Leading Order results do not differ appreciably.

\noindent iii) The ratio $F_2^n/F_2^p$ differs from the value 2/3, the 
asymptotic
value for $SU(6)$-symmetric models, because of the presence of spin-dependent 
forces in the model Hamiltonian.

\noindent iv) Both the chromomagnetic (hyperfine) interaction of the IK model 
and the chiral mechanism of the GBE model fail to reproduce
the correct behaviour of the experimental data in the region of validity of 
our calculation, namely $0.3 \lesssim x \lesssim 0.7$. 
A conclusion already known in the case of the IK model \cite{noi1}, but valid 
also for the GBE model.
Note that this discrepancy is highly significant, since it 
concerns the region of Bjorken-$x$ where valence quarks dominate. Therefore
complications arising from sea and gluon contributions cannot affect 
these results.

\noindent v) 
The effects of the large amount of high-momentum components generated by the
relativistic expression of the kinetic energy can be noticed in the case
of the GBE model, where the deviation with respect to the   
$SU(6)$-symmetric value 2/3 is shifted at higher value of $x$.
Similar effects are found in a calculation \cite{simu} which 
makes use of the relativized extension of the IK model
proposed by Isgur and Capstick\cite{ICa}.

\noindent vi) It should be stressed that the disagreement emerges only if the 
Pauli principle effects are properly included. A simple model of the scattering
 involving separately the effects from an active quark plus a core of spin 1 
and 0 quark pairs, would naturally reproduce the decreasing behaviour of the 
ratio $F_2^n/F_2^p$ as function of $x$\cite{close73,carlitz75,closethomas88}.

\noindent vii) The disagreement with the 
experimental data does not depend on the details of the wave-function, 
but on the specific form of the $SU(6)$ breaking terms
introduced in the model Hamiltonian. They are quite similar in the
case of GBE and OGE models\cite{canotra} despite of the different 
dynamical mechanisms invoked.

\vspace{1mm}

In conclusion, by means of a direct and transparent light-front 
calculation, we have shown that the ratio between the neutron and the proton 
structure 
functions is sensitive to the spin-dependent part of the quark-model 
Hamiltonian describing $SU(6)$-breaking effects at low hadronic energy. The 
correct theoretical behaviour
of the ratio can be obtained only if the Pauli principle is preserved in the 
explicit calculation of the structure functions. Both hyperfine 
interactions, from the one-gluon-exchange potential model and from the spin 
dependent part of the Goldstone-boson interaction, cannot reproduce the 
large-$x$ behaviour of the data even at the qualitative level. The 
$SU(6)$-breaking mechanism seen in deep inelastic scattering seems to be 
different from the dynamical effects at low-energy, at least in the two cases 
we studied.
Previous studies of unpolarized structure functions \cite{noi1} show 
that a model able to invert the tendency of a ratio larger than 2/3 in the 
large-$x$ region is the algebraic model proposed by Iachello \cite{algebra}. 
Despite the fact that the model is built in such a way that only the 
symmetry properties are retained and no dynamical mechanisms can be deduced, 
the parametrization which emerges is consistent with the data, in particular 
if the structure of the (effective) constituent quark is taken into account 
\cite{SVT97ss,ScoVe01}.
Work to include the instanton dynamics and to enlarge the study of
$SU(6)$ breaking effects into the region of polarized parton
distributions is in progress.

\section*{Acknowledgments} 
We acknowledge useful conversations with
Sergio Scopetta, Silvano Simula, and Vicente Vento. 
We are grateful to Wolfram Weise 
for a careful reading of the manuscript and interesting comments.  
We also thank R.F. Wagenbrunn for providing us with the nucleon wave-functions
in the GBE model.

\newpage

\begin{center}

{\bf Figure captions}

\end{center}

\vspace{5mm}

\noindent Fig.~1~a: The proton structure function as predicted by the
IK and GBE models at the hadronic scale (dotted and dot-dashed curves, 
respectively), and at the scale $Q^2 = 10$ GeV$^2$ 
(dashed and continuous lines, respectively).
Fit to the experimental data from the CTEQ5(D) 
analysis of ref.~\cite{cteq5}: triangles.

\vspace{1mm}

\noindent Fig.~1~b: As in Fig.~1~a for the neutron structure function.

\vspace{20mm}

\noindent Fig.~2~a: The ratio $F_2^n/F_2^p$ as function of $x$ for the GBE 
model at the hadronic scale (dot-dashed curve), and at the scale $Q^2 = 10$ 
GeV$^2$; Leading-Order evolution (dotted curve), Next-to-Leading-Order 
evolution (full curve). Fit to the experimental data from the CTEQ5(D) 
analysis of ref.~\cite{cteq5}: triangles.

\vspace{1mm}

\noindent Fig.~2~b: The ratio $F_2^n/F_2^p$ as function of $x$ for the IK 
model. Notations as in Fig.~2~a.

\newpage

\protect

\begin{figure}[h]
\begin{center}
\mbox{
\epsfig{file=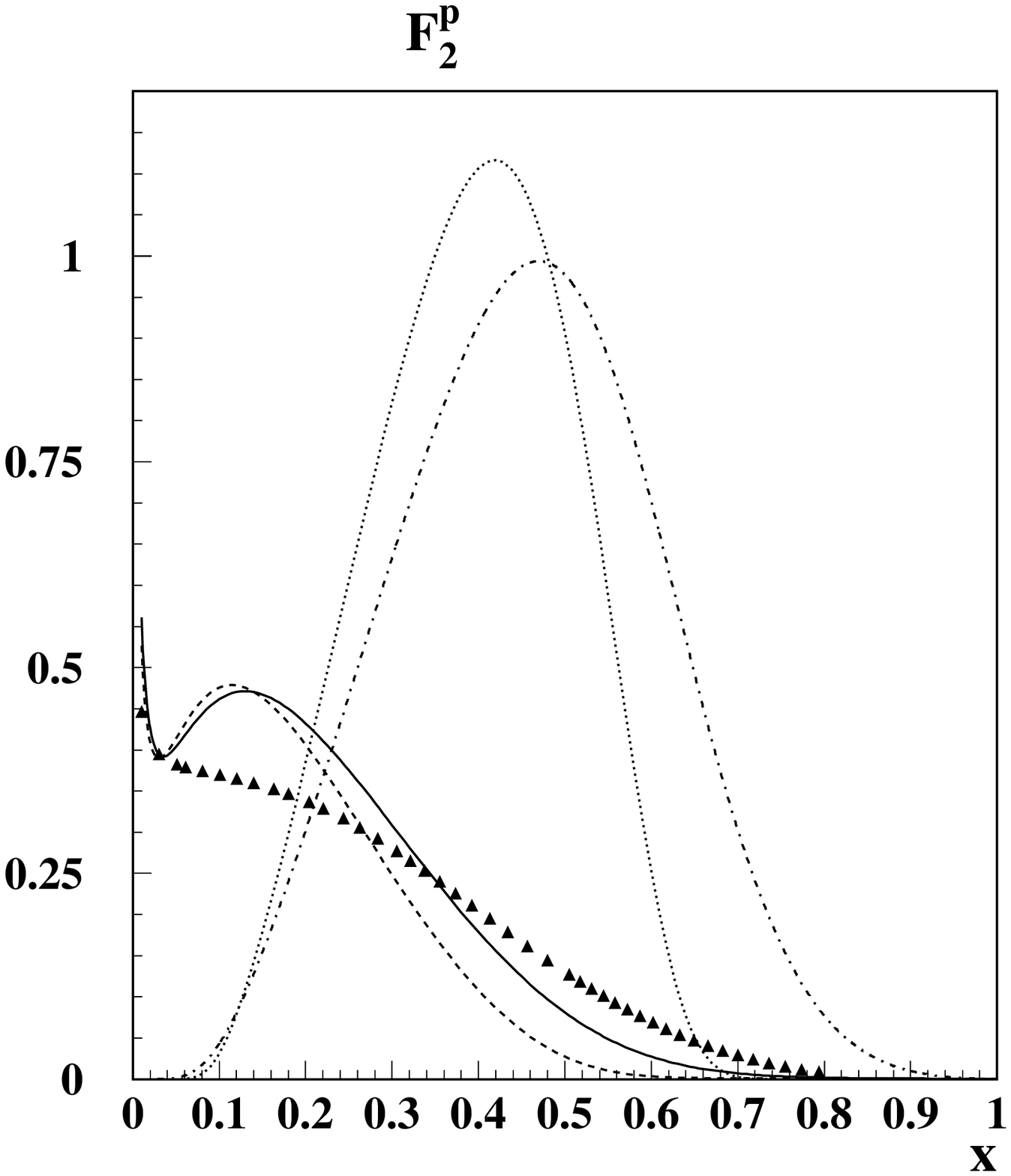, width =11cm, height=10. cm}}
\end{center}
\vspace{-5 truemm}
\centerline{\bf \large Figure 1~a}
\begin{center}
\mbox{
\epsfig{file=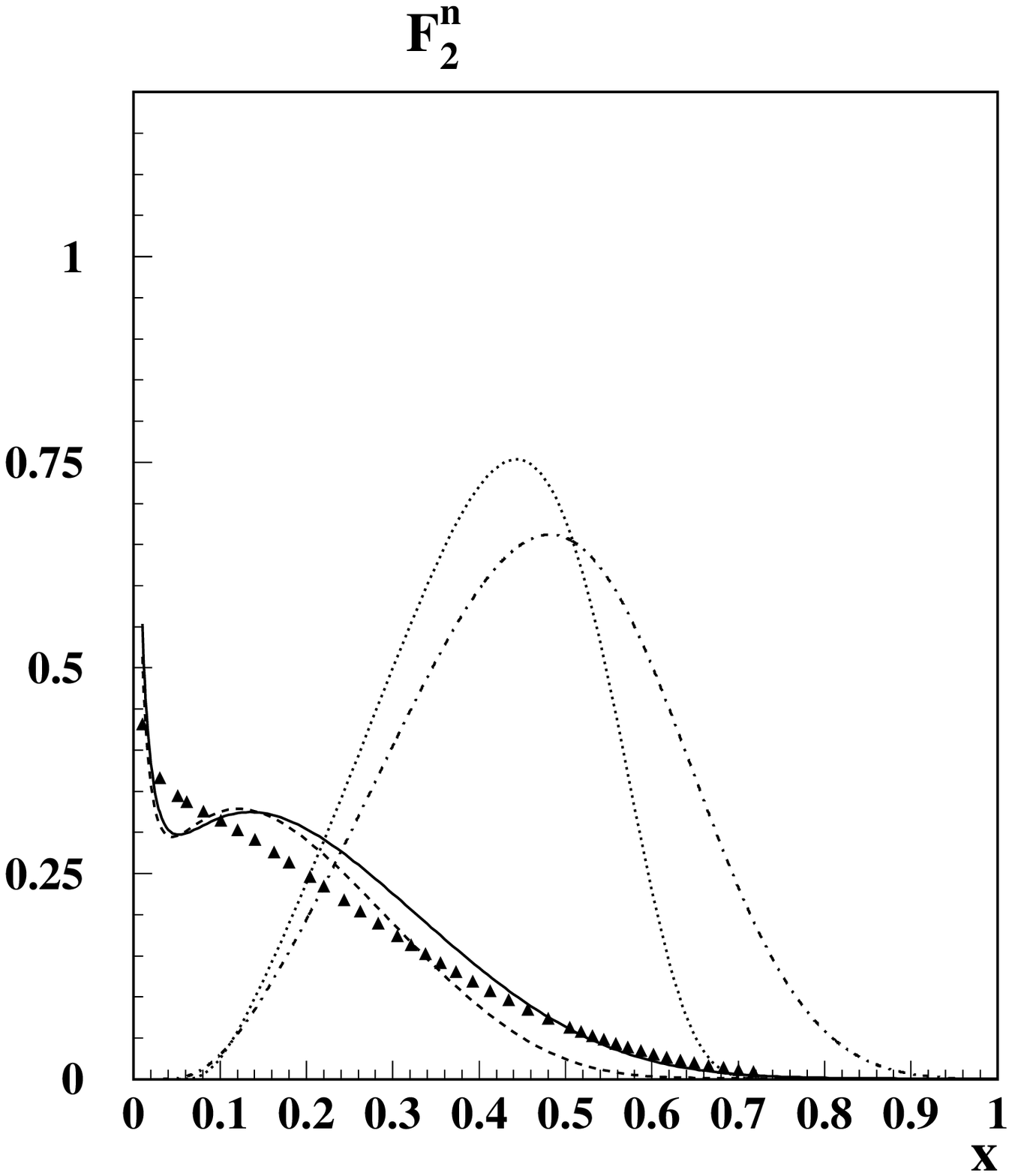, width=11cm, height=10. cm}}
\end{center}
\vspace{-5 truemm}
\centerline{\bf \large Figure 1~b}
\end{figure}

\newpage

\begin{figure}[h]
\begin{center}
\mbox{
\epsfig{file=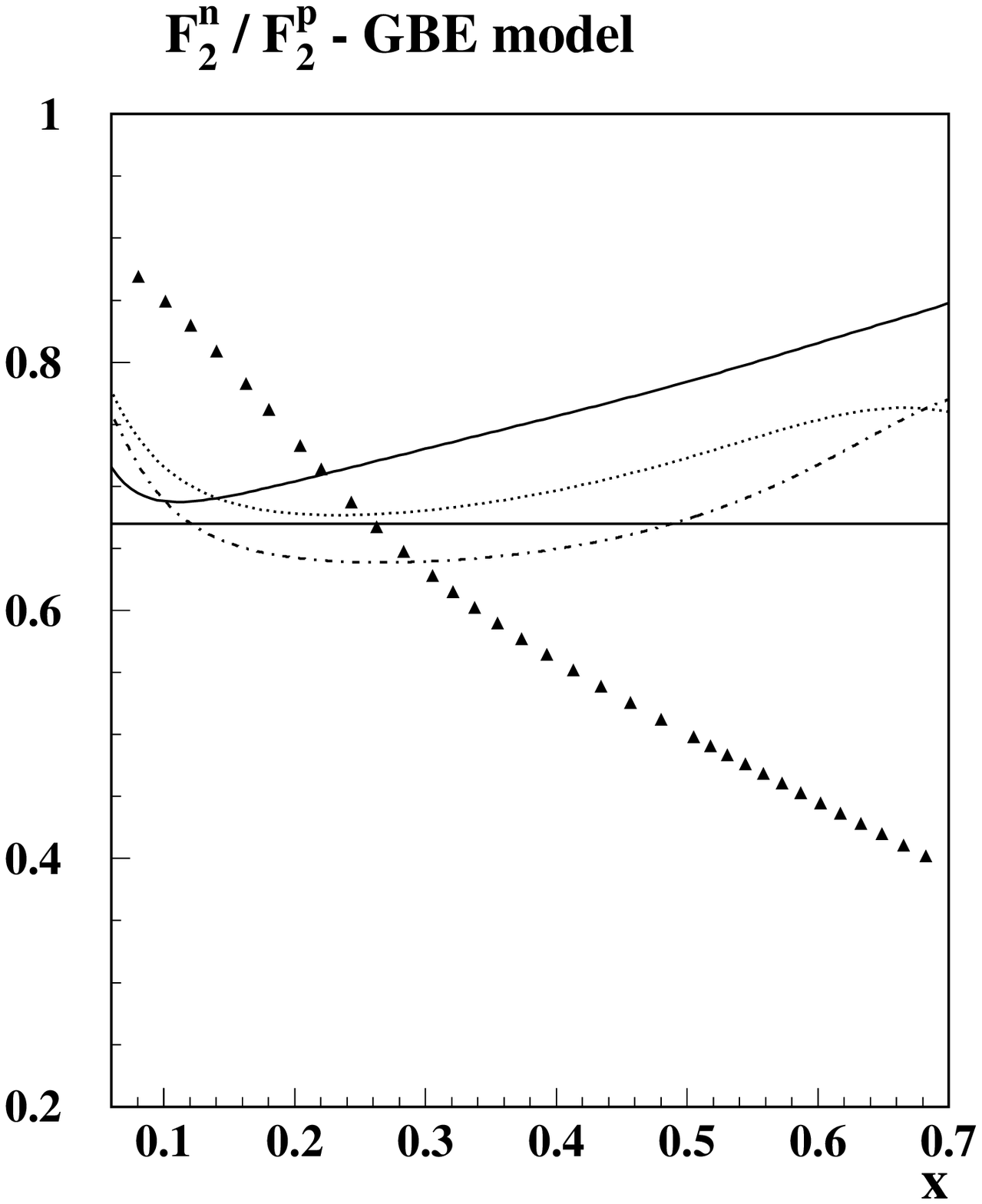, width =10 cm, height=10. cm}}
\end{center}
\vspace{-5 truemm}
\centerline{\bf \large Figure 2~a}
\begin{center}
\mbox{
\epsfig{file=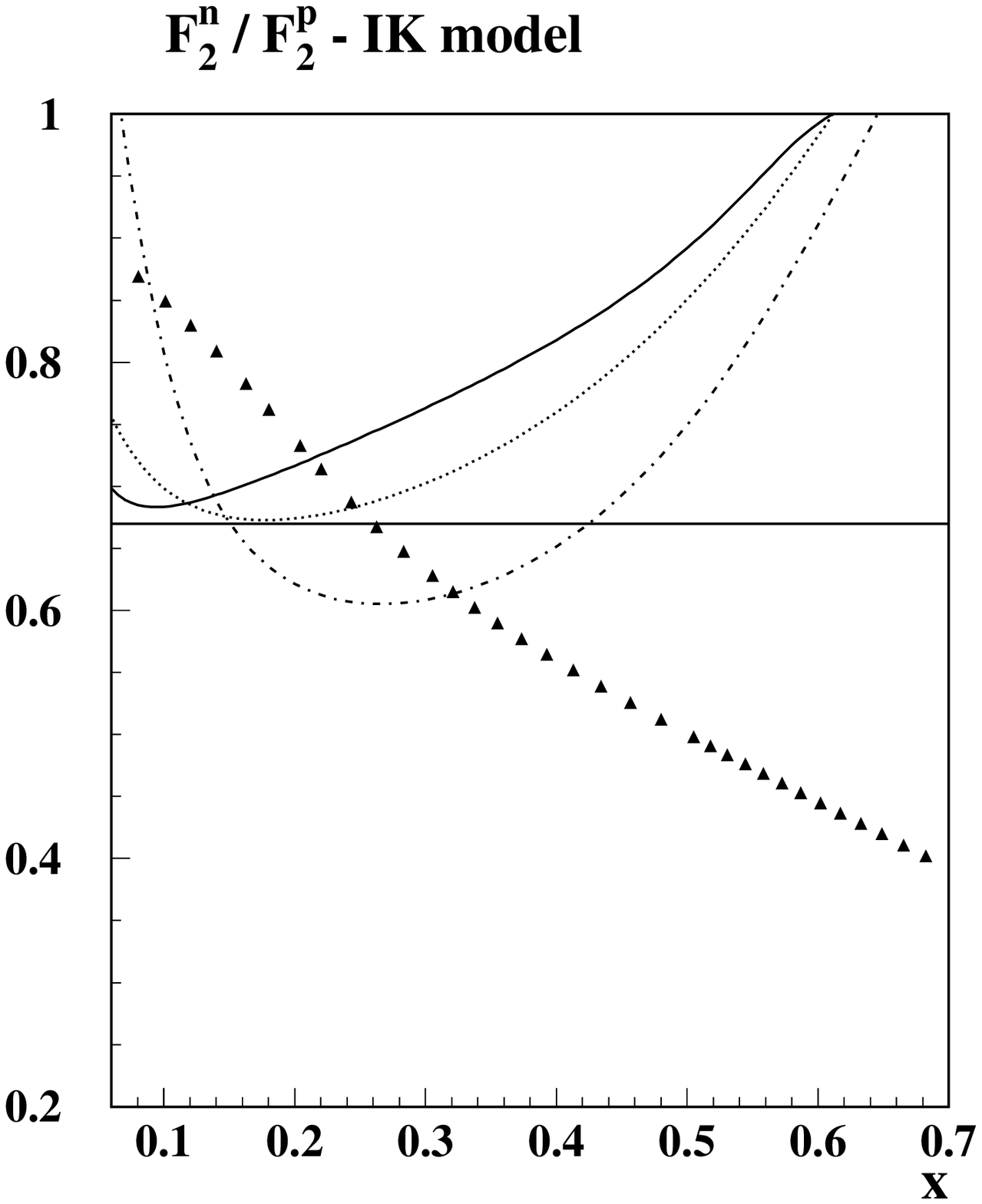, width=11 cm , height=10. cm}}
\end{center}
\vspace{-5 truemm}
\centerline{\bf \large Figure 2~b}
\end{figure}

\end{document}